\documentstyle[epsfig,preprint,aps,tighten]{revtex}

\begin{document}

\draft

\title{Effective Baryon-Baryon Potentials in the Quark Delocalization and
       Color Screening Model}

\author{Jia-Lun Ping}
\address{ Department of Physics, Nanjing Normal University, Nanjing,
 210097, P.\ R.\ China}
\author{ Fan Wang}
\address {Center for Theoretical Physics and Department of Physics,
	  Nanjing University, Nanjing, 210008, P.\ R.\ China.}
\author{ T.\ Goldman}
\address {Theoretical Division, Los Alamos National Laboratory,
 Los Alamas, NM 87545, U.S.A }

\date{\today}
\maketitle

\vspace{-4.0in}
\flushright{LA-UR-98-5727}
\vspace{4.0in}

\begin{abstract}
The quark delocalization and color screening model is used for a
systematic study of the effective potential between baryons in the
$u,d$ and $s$ sector. The model is constrained by the properties of
baryons and $N$-$N$ scattering. The effective potentials for the
$N$-$N$ ($IJ=01,10,11,00$) channels and the $N$-$\Lambda$ and
$N$-$\Sigma$ ($IJ = \frac{1}{2}1, \frac{1}{2}0, \frac{3}{2}1,
\frac{3}{2}0$) channels fit the $N$-$N$, $N$-$\Lambda$ and $N$-$\Sigma$
scattering data reasonably well. This model predicts:  There are rather
strong effective attractions between decuplet-baryons; the effective
attractions between octet-baryons are weak or even repulsive; and the
attractions between decuplet- and octet-baryons lie in between.
\end{abstract}

\pacs{25.80.-e 25.10.+s 13.75.-n 25.40.-h}
\flushleft

\newpage
\section{Introduction}

Realistic models for baryon-baryon interactions are relevant to the
study of the strong interaction, the understanding of properties of
nuclei and hypernuclei, multiquark states, neutron stars, stangelets,
strange matter and so on.  A simple but realistic baryon-baryon
potential is the basic input for any realistic calculation of
hypernuclear properties and the equation of state of strongly
interacting matter\cite{hyp,Dover,eos}.  Such a potential is also
needed for the study of dibaryons, as it is clear that an effective
attraction between two baryons is necessary for the formation of a
dibaryon, although it is not sufficient.

There are several approaches to obtaining the effective potential
between baryons. One-boson-exchange-models, such as the Nijmegen and
J\"ulich models\cite{obe}, (although there are many others,) are a
typical approach.  The advantage of this approach is that it gives a
very accurate description of the nucleon-nucleon (NN) interaction. For
extension to the hyperon-nucleon (YN) interaction, the problem of too
many unkown parameters (coupling constants) can be solved by imposing
$SU_3$ flavor symmetry to constrain the coupling constants. The
disadvantage is that a phenomenologically repulsive core, which is
channel-dependent, has to be used.

Another approach\cite{Oka,Straub,Fujiwara} is based on applying the
resonating group method (RGM) of calculation to quarks. Here, however,
gluon exchange between the constituent quarks can only give the short
range, repulsive core for the NN interaction. So in this model, meson
exchange must be introduced to reproduce the medium and long range part
of baryon-baryon interactions. It does have the advantage that the
short range part is described by the quark model instead of a
phenomenological repulsive core.

Although the experimental data can put some constraints on models of
baryon-baryon interactions, the existing data, even including the
newest data, still can not discriminate among most models. This is true
because any disagreements with the newer data can generally be repaired
by fine-tuning parameters, relaxing some of the constraints or making
other adjustments in the models\cite{tim}.  So in order to have a
meaningful model for the baryon-baryon interaction with some predictive
power, a model which is ``parameter-free'' is highly desirable.

Recently a new quark model, the quark delocalization, color screening
model, (QDCSM)\cite{prl}, has been developed. On comparison with the
conventional quark cluster model, it is apparent that two new
ingredients have been introduced. One is that the Hilbert space is
enlarged to include fully confined and fully deconfined configurations
as two extremes so that the actual configuration is determined
variationally by the dynamics of the six quark system. The other is
that the possible difference of the $q-q$ interaction inside a hadron
(in an overall colorless state) and between quarks from two different
colorless hadrons, due to the nonlinearity of QCD\footnote{For example,
the contribution from gluons exchanged among three quarks which are in
a color-singlet state is zero, while there is a nonzero contribution
among three quarks which are in "colorful" state.}, is taken into
consideration.

In this way, without introducing mesons, the long standing problem of
the missing medium range attraction of NN forces in quark models is
solved, and the parallel between nuclear and molecular forces obtains a
natural explanation. The model has been successfully applied to
describe NN scattering\cite{prl,QDCSM}.  Using this model, a systematic
search for dibaryon candidates within the  three flavor world of
$u,d,s$ quarks was also made\cite{DB}. A relativistic model calculation
obtains qualitatively similar results, even though the confinement
mechanism differs.~\cite{sysreldib}  Therefore, it seems worthwhile to
apply this model to perform systematic and parameter free calculations
of the effective baryon-baryon potentials.

This paper is organized as follows: In section II, the model
Hamiltonian and Hilbert space are described. Section III is devoted to
a sketch of the calculation method. The results are given in section IV
and a conclusion is given in section V.

\section{Quark Delocalization, Color Screening Model}

The details of the model can be obtained from ref.~\cite{QDCSM,DB}. Here
we only give a brief review.

For a single hadron, the QDCSM is same as the quark potential model,
which has been successfully applied to describe the properties of
baryons and mesons.  The model Hamiltonian which describes a single baryon
is written as
\begin{eqnarray}
H(3) & = & \sum_{i=1}^3 (m_i+\frac{p_i^2}{2m_i}) +\sum_{i<j=1}^{3}V_{ij} 
	  -T_c(3),   \\
V_{ij} & = & V_{ij}^c + V_{ij}^G, \nonumber \\
V_{ij}^c & = & -{\vec \lambda_i}\cdot {\vec \lambda_j}ar_{ij}^n, ~~~n=1,2,
  \nonumber \\
V_{ij}^G & = & \alpha_s \frac{ {\vec \lambda_i} \cdot {\vec \lambda_j} }{4}
 \left[ \frac{1}{r_{ij}}-\frac{\pi}{2} \left( \frac{1}{m_i^2}+\frac{1}{m_j^2}
 +\frac{4}{3m_im_j} {\vec \sigma_i} \cdot {\vec \sigma_j} \right) 
 \delta (\vec{r}_{ij}) + \cdots \right]. \nonumber
\end{eqnarray}
The $m_i$, $p_i$ and $r_{ij}$ are quark masses, momenta and pairwise
separations. The ${\vec \lambda_i}$ are color SU(3) Gell-Mann matrices
and $\alpha_s$ is the strong coupling constant. For the confinement
potential, $V^c$, both linear and quadratic forms are used in our
calculations.  In the effective one gluon exchange potential, $V^G$,
only the color Coulomb and color magnetic terms are retained; other
terms have been neglected temporarily {\cite{QDCSM}} in order to reduce
the calculational burden.

The wave function (WF) for a single baryon has the form 
\begin{equation}
\psi (123)=\chi (123)\eta _{SIJ}(123)\phi (123). \label{wf3}
\end{equation}
Here $\chi (123)$ is the singlet color WF, $\eta _{SIJ}(123)$ is the
symmetric spin-flavor SU$_{2\times f}^{f\sigma }\;\supset $SU$_f\times
$SU$_2^\sigma $ WF (with $S=$strangeness, $I=$isospin, $J=$spin). The
spatial wavefunction is given by
\begin{eqnarray}
\phi (123) & = & \phi _{1s}({\vec r_1})\phi _{1s}({\vec r_2})
	   \phi _{1s}({\vec r_3}),  \nonumber \\
\phi _{1s}({\vec r}) & = &  \left( \frac{1}{\pi b^2} \right)^{\frac{3}{4}}  
		     e^{-\frac{({\vec r}-{\vec s})^2}{2b^2}}, \label{gs} 
\end{eqnarray}
where ${\vec s}$ is a reference center and $b$ is a baryon size
parameter.

For the two baryon system, we extend the Hamiltonian to
\begin{equation}
H(6) = \sum_{i=1}^6 (m_i+\frac{p_i^2}{2m_i}) +\sum_{i<j=1}^{6}V_{ij}
          -T_c(6), \label{ham}
\end{equation}
but modify the color confinement as follows:
\begin{equation}
V_{ij}^c=\left\{ 
\begin{array}{ll}
-\vec{\lambda}_i\cdot {\vec \lambda}_j ar_{ij}^n & ~~~ 
\mbox{if }i,j\mbox{ occur in the same baryon orbit}, \\ 
-{\vec \lambda}_i \cdot {\vec \lambda}_j \frac{a}{\mu_n}(1-e^{-\mu_n
r_{ij}^n}) & ~~~ \mbox{if }i,j\mbox{ occur in different baryon orbit}. 
\end{array}
\right. 
\end{equation}
Screened linear color confinement is found in lattice QCD
calculations\cite{LGQCD} and can be parametrized as
\begin{equation}
V(r)=\left( -\frac{\alpha _s}r+\sigma r\right) \left( \frac{1-e^{-\mu_1 r}}{%
\mu_1 r}\right), 
\end{equation}
with $\alpha _s=0.21\pm 0.01,~~~ \sqrt{\sigma }=400\,\mbox{MeV}, ~~~ \mu_1
^{-1}=0.90\pm 0.20\,\mbox{fm}. $

For the two baryon wave function, we extend the quark cluster model
space by introducing delocalized single quark orbits:
\begin{eqnarray}
\psi_L({\vec r}) & = & \left( \phi_L({\vec r})+\epsilon(s) \phi_R({\vec r})
\right)/N(s), \nonumber \\
\psi_R({\vec r}) & = & \left( \phi_R({\vec r})+\epsilon(s) \phi_L({\vec r})
\right)/N(s), \label{dl} \\
N^2(s) & = & 1+\epsilon^2 (s)+2\epsilon (s)\langle\phi_L|\phi_R\rangle.
  \nonumber
\end{eqnarray}
where $\phi_L,\phi_R$ are the quark cluster bases of the type described
above for individual baryons:
\begin{equation}
\begin{array}{ll}
\phi_L({\vec r}) = \left(\frac{1}{\pi b^2} \right)^{\frac{3}{4}} 
  e^{-\frac{({\vec r}+{\vec s}/2)^2}{2b^2}} ~~~~ & 
  \mbox{(left centered orbit)}, \\ 
\phi_R({\vec r}) = \left(\frac{1}{\pi b^2} \right)^{\frac{3}{4}} 
  e^{- \frac{({\vec r}-{\vec s}/2)^2}{2b^2}} ~~~~ & 
  \mbox{(right centered orbit)}. \label{orbit}
\end{array}
\end{equation}
Here ${\vec s}$\  is the separation between the centers of two $q^3$
clusters.  The delocalization parameter $\epsilon (s)$ is determined
variationally for every ${s = |\vec s|}$ by the $q^6$ dynamics (see
section III). The quark molecular orbit introduced by Fl. Stancu and
L.  Wilets\cite{SW} is a restricted version of this structure.

The two baryon wave function is written as
\begin{eqnarray}
\Psi^{\alpha}_{\alpha_1 F_1,\alpha_2 F_2} (1\cdots 6) & = & {\cal A} 
[\psi_{\alpha_1 F_1}(123)\psi_{\alpha_2 F_2}(456)]_{\alpha}, \nonumber \\
\psi_{\alpha_1 F_1}(123) & = & \chi (123) \eta_{S_1I_1J_1F_1}(123)\psi_L (1)
\psi_L (2)\psi_L (3) \label{basis} \\
\psi_{\alpha_2 F_2}(456) & = & \chi (456) \eta_{S_2I_2J_2F_2}(456)\psi_R (4)
\psi_R (5)\psi_R (6). \nonumber
\end{eqnarray}
Here $\alpha =(SIJ)$ are again the strong interaction conserved quantum
numbers:  strangeness, isospin and spin. (Orbital angular momentum is
assumed to be zero for the lowest states. In principle, an angular
momentum projection should be done but we leave this for future
refinement.) The $q^3$ cluster WF is almost the same as given in
Eq.(\ref{wf3}), but the single cluster Gaussian WF, Eq.(\ref{gs}), is
replaced by the delocalized orbital WF, eq.(\ref{dl}). Finally, ${\cal
A}$ is the normalized antisymmetry operator
$$
{\cal A}=\frac{1}{\sqrt{20} }\sum (-)^{\delta _p}p 
$$

The model parameters $m, m_s, b, \alpha_s, a $ are fixed by the
spectrum of baryons; $\mu_1$ is taken from lattice gauge QCD
calculations; and $\mu_2$ is fixed by fitting the $N$-$N$ phase shifts.
The fitted parameters are given in Table I.

\begin{center}
Table I. The parameters used in the calculations.

\begin{tabular}{|l|c|c|c|c|c|c|c|c|} \hline
       & $m$(MeV) & $m_s$(MeV) & $b$(fm) & $\alpha_s$ 
       & $a$(MeV$\cdot$fm$^{-2})$ & $\mu_n$  \\ \hline 
 $n=1$ &  313     & 618.45     & 0.625   & 1.71    & 39.14 & 1.1  \\ \hline 
 $n=2$ &  313     & 633.76     & 0.603   & 1.54    & 25.13 & 1.6  \\ \hline 
\end{tabular}
\end{center}

\section{Calculation method}

Since the relative motion between two baryons is rather slow compared
to the internal motion of the quarks in the baryons and the quantum
fluctuation of the center of mass coordinates of the baryons is
neglected temporarily (again leaving refinement for the future), we can
define the effective baryon-baryon potential  as follows:
\begin{equation}
V_{\alpha}(s_0) = E_6(s=s_0) - E_6 (s=\infty) \label{e61}
\end{equation}
where $E_6(s)$ is the energy of two baryon system at separation
$s$, which can be obtained by calculating the matrix elements of the
Hamiltonian, Eq.(\ref{ham}), in the baryon states, Eq.(\ref{basis}). In
the nonrelativistic approximation,
\begin{equation}
 E_6(s) = E_3(123) +E_3(456) + E_{rel} + V_{\alpha}(s), \label{e62}
\end{equation}
where $E_3$ is the total energy of a single baryon and $E_{rel}$ is the
energy of relative motion of two baryons.  When $s \rightarrow \infty$,
we must have $V_{\alpha} \rightarrow 0$, so
\begin{equation}
 E_6(s=\infty) = E_3(123) +E_3(456) + E_{rel} . \label{e63}
\end{equation}
Note that the six-quark WF decomposition maintains a non-zero $E_{rel}$
even at $\infty$ and it is our assumption that this constant is the
appropriate quantity to remove at finite $s$ which leads to the
expression for $V_{\alpha}(s)$ shown in Eq.(\ref{e61}).

The treatment of the center of mass motion merits some additional
discussion.  In order to have a correct separation of total kinetic
energy into internal, relative and center of mass parts in the
asymptotic region, we redefined the CM energy using an average of the
single quark mass
\begin{equation}
T_c(n) = \frac{1}{2M_n} \left( \sum_{i=1}^{n} \vec{p}_i \right)^2
    = \frac{1}{2} \left( \frac{n-n_s}{n} \frac{1}{m} +\frac{n_s}{n} 
      \frac{1}{m_s} \right) \left( \sum_{i=1}^{n} \vec{p}_i \right)^2 ,
\end{equation}
where $n$ and $n_s$ are total number of quarks and $s$-quarks in the
system, respectively. In this way, the kinetic energy is free from CM
motion when an SU(3)-flavor symmetric wave function is used.
\begin{equation}
T(n) = \sum_{i=1}^{n} \frac{1}{2m_i} {\vec{p}_i}^2 - T_c(n)
  = \frac{1}{2nm} \left( \frac{n-n_s}{n}+\frac{n_s}{n} \frac{m}{m_s}
     \right) \sum_{i>j=1}^{n} ( \vec{p}_i -\vec{p}_j )^2
\end{equation}
A similar treatment can be found in ref.~\cite{Fujiwara}.

To calculate the matrix elements of $H$ in the basis given by
Eq.(\ref{basis}) is a daunting task, especially for a systematic
search, where many matrix elements must be evaluated. So a group theory
method was developed to relate the physical basis to a group chain
classification basis (symmetry basis). In the symmetry basis, the well
known fractional parentage expansion can be used. The calculation of
six-body matrix elements reduces to two-body matrix elements and
four-body overlaps. For details of the method, see ref.~\cite{GT}.

In order to discern the effect of channel coupling, a channel potential
is introduced.  For a given set of quantum numbers $\alpha =(SIJ)$, the
effective channel potentials are defined by the same formula
Eq.(\ref{e61}), where $E_6(s=s_0)$ and $E_6(s=\infty)$ are the 
eigenenergies of the two baryon system at separation $s_0$ and
$\infty$, respectively, which can be obtained by diagonalizing the
$q^6$ Hamiltonian. The eigenstates are expressed as multiple physical
channel coupling wave functions
\begin{equation}
\Psi _\alpha (1\cdots 6)=\sum_{\alpha _1F_1,\alpha _2F_2}C_{\alpha
_1F_1,\alpha _2F_2}^\alpha \Psi _{\alpha _1F_1,\alpha _2F_2}^\alpha . 
\end{equation}
The channel coupling coefficients $C_{\alpha _1F_1,\alpha _2F_2}^\alpha$ 
are determined by the diagonalization. 

In general, we can identify which of the baryon channels an eigenenergy
corresponds to by finding which baryon channel for that eigenenergy has
the largest channel coupling coefficient. Although this identification
is not unique, it guarantees that the channel energy varies smoothly
with $s_0$. Where no coefficient is discernably dominant, we examine
the energies at (nearby) larger and smaller values of $s_0$ and again
choose the value which provides for a smooth continuation of the
channel energy. This replaces an earlier procedure we used which
involved choosing the lowest eigenenergy available. This older
procedure had the disadvantage of not having a clear relation to
particular baryon-baryon channels, of introducing sharp (discontinuous
in the derivative with respect to $s_0$) changes in the energy, and of
prohibiting the identification of any but the lowest energy channel.

The energies obtained in this way depend on the separation ${s}$ and
the delocalization parameter $\epsilon (s)$. We repeat the calculation
for each ${s}$ by varying $\epsilon (s)$ from 0.01-1.0 with 0.01
stepsize to obtain the minimum of the lowest eigenenergy. At the same
time, the delocalization parameter $\epsilon (s)$ and all eigenenergies
are determined.

\section{Results}
 
The effective potential between combinations of octet- and
decuplet-baryons (168 channels, altogether) were calculated. To save
space, only a few typical examples are shown in Figs.1--7. (We will be
happy to provide any of the others upon request.)

Fig.1a shows the effective $N$-$N$ potential for channels (IJ)=(01),
(10), (11) and (00) with the quadratic color confinement interaction.
Clearly there is an effective attraction in the first two channels,
with (01) a little stronger than (10). The other two channels are 
repulsive.  These potentials are quite similar to phenomenological
$N$-$N$ potentials such as the Reid soft core $N$-$N$
potential.~\cite{Reid} However, when we proceed to a dynamical
calculation of the phase shifts, we must also perform a partial wave
decomposition and make a center of mass correction of the kernel of the
interaction Hamiltonian. The net effect of these calculations is to
create some additional attraction, so that a slight reduction of the
value of the screening parameter is then needed to produce good
agreement with the results of $N$-$N$ phase shift analyses.~\cite{rmpla}

Fig.1b is similar to Fig.1a but with the linear color confinement
interaction. Comparison of Fig.1a with Fig.1b makes it clear that both
sets of potentials have qualitatively the same pattern, but with a
weaker attraction for the case of linear confinement. In fact, we find
that for the linear color confinement interaction, the potentials in
these four channels agree reasonably well~\cite{prep} with the phase
shift analyses {\em without any change from the lattice QCD value of
the screening parameter}~\cite{LGQCD}. In this sense, our results
provide a parameter-free prediction of the phase shifts in these
channels.

The channel coupling calculation shows that channel coupling produces
minor effects for the $N$-$N$ channels, and so we do not present those
results here.   In the following, we only present the results with
quadratic color confinement to save space.

Figs.2-3 shows the results for $N$-$\Sigma$ and $N$-$\Lambda$
channels.  For $N$-$\Lambda$, the spin triplet state is a little more
attractive than the spin singlet; the channel coupling adds a bit
more attraction to this state but leaves the spin singlet almost
unchanged.  We note that $^{4}_{\Lambda}$H and $^{4}_{\Lambda}$He both
have spin zero ground states and spin one excited states. One might
interpret this as evidence that the spin-singlet $N$-$\Lambda$
interaction is more attractive than the spin triplet. However, the
situation is not so transparent due to the complications of the
interactions of the four bodies involved and, in addition,
$\Lambda$-$\Sigma^{0}$ mixing effects. The spin one ground state of the
deuteron is certainly an indication that the spin triplet $N$-$N$
interaction is more attractive than the spin triplet, and we might
reasonably expect (by flavor symmetry) that this should hold true for
all octet-baryon combinations. However, due to the strong tensor
interaction from pion exchange, the Nijmegen OBE model F~\cite{obe}
nonetheless includes a more attractive spin singlet $N$-$N$ central
interaction. Furthermore, there is a paucity of direct data on
scattering in the $Y$-$N$ channels, and widely differing relative
strengths for the central interaction are all consistent with both the
available data and the nuclear states referred to above. The QDCSM, on
the other hand, predicts that spin triplet $N$-$N$ and $N$-$\Lambda$
interactions are stronger than spin singlet ones.  Clearly, which
central interaction is stronger in each case merits additional study.

For $N$-$\Sigma$, we find the strongest attraction in the
$IJ=\frac{1}{2}1$ channel, while the $IJ=\frac{3}{2}0$, and
$\frac{1}{2}0$ channels both have a little weaker attraction (single
channel case), and the $IJ=\frac{3}{2}1$ channel is repulsive.  Channel
coupling has little effect on $\frac{1}{2}1$, and pushes $\frac{1}{2}0$
from attractive to repulsive. These show that the $N$-$\Sigma$
potential is more strongly spin and/or isospin dependent than the
$N$-$\Lambda$ potentials, which have a little weaker dependence on
spin.  These results are in qualtitative agreement with the
calculations of OBE models~\cite{obe} and hybrid quark model
calculations~\cite{Straub}, except that our attraction for the
$N$-$\Sigma$($\frac{1}{2}1$) channel is too strong.  It is worth noting
that the $N$-$\Sigma$ ($\frac{3}{2}1$) channel has a small bump at $R_s
\sim 1.3$fm. The reason for this is not clear, but it seems to
correspond to the statement of ref.~\cite{Straub}, that the repulsion
between two clusters has its maximum not at distance zero, but at a
finite cluster distance.

For all other channels, some general features of the effective
potential have been found in this calculation:  An effective attraction
exists in most channels. The attraction is very strong between two
decuplet-baryons, but the attraction is very weak and even repulsive
between octet-baryons, while the strength of the attraction between
octet- and decuplet-baryons lies in between. The reason is that in a
decuplet baryon, spin is totally symmetric and color is totally
antisymmetric, so that the hyperfine interaction is purely repulsive.
For a six quark system, the spin and color are neither totally
symmetric nor antisymmetric, leading to the hyperfine interaction not
always being repulsive, and so providing an effective attraction.

Some states are forbidden for structureless baryons by the Pauli
principle: $\Delta\Delta(IJ=33)$, $\Delta\Sigma^{*}(\frac{5}{2}3)$,
$\Sigma^*\Xi^*(\frac{1}{2}3)$, $\Omega\Omega(03)$.  For these states,
we find that the energy of the system becomes very large when the
distance between the two baryons tends to zero (see Fig.4).

Because the effective $B$-$B$ interactions are directly related to
multiquark states, in the following we present the baryon-baryon
potential along with a discussion regarding dibaryon states.

(a) States with only non-strange quarks. Since the states in this part
do not contain any strange quarks, they are not affected by symmetry
breaking.

Clearly many channels have an effective attraction, but to form bound
states one has to consider not only two body decay modes, but also many
body decay modes. For example, for a $\Delta \Delta$ channel to form a
narrow dibaryon resonance, the energy of dibaryon should not only be
less than the sum of the masses of two $\Delta$s, but it should also be
less than the sum of the masses of two nucleons and two pions because
each $\Delta$ itself can decay into $N\pi$. In addition, the zero-point
harmonic oscillater energy must be added to the system.  Only channels
with sufficient attraction can form bound states or narrow resonances.

The interesting states are:\\ 
$NN(01)$, $NN(10)$ and $\Delta \Delta (03)$. The $NN(01)$ state is a
loosely bound state: This is just the deuteron, which is reproduced
here. The $NN(10)$ state corresponds to the known $NN$ zero energy
resonance.  The $\Delta \Delta (03)$, called $d^*$, is a narrow
resonance.~\cite{fanetal} Although the $\Delta \Delta (01)$ channel is
more attractive than the $\Delta \Delta (03)$ (see Fig.5), it couples
strongly to $NN(01)$, so it is not a good candidate for an observable
dibaryon. Conversely, because the $\Delta \Delta (03)$ has a large
angular momentum, it cannot couple to an $NN$ $s$-wave state, or to
$N\Delta$ (also forbidden by isospin) directly. The large angular
momentum hinders the decay to $NN$, and the energy of the state
prevents its decay to $NN\pi\pi$, so it is a good candidate for a
narrow resonance.~\cite{cwong}

(b) States with one or more strange quarks. The general features stay
the same for states with strange quarks.  There are also several
particularly interesting states (see Figs.6-8).

$\Sigma\Sigma (20), ~\Xi\Xi (00)$ are both deuteron-like states,
consisting of two octet-baryons, which have a weak attraction and the
minima of their potentials are found at rather large separations ($>1$
fm). Channel coupling has a minor effect on the results.  These are
possible dibaryon candidates but are sensitive to model details.

It is worth mentioning the $\Omega\Omega (00)$ state. In our
calculations, its effective attraction is not weak, but rather strong
enough to form a strong interaction bound state.  Because it consists
of two $\Omega$s and $\Omega$ is strong interaction stable, it is a
good candidate for a dibaryon, although it is also sensitive to model
details. In addition, it may be too difficult to produce this state
even in relativistic heavy ion collision experiments.

The $H$ particle ($YIJ=000$) is a special state in our calculation. The
channel coupling calculation (see Fig.6) shows that it is a six quark
state consisting mainly of octet-baryons (dominant components are
$N\Xi$ and $\Sigma\Sigma$) that has an attraction that is not too weak.
It is possible that it appears as a dibaryon.

The states $YIJ=-1\frac{1}{2}2$ and $-1\frac{1}{2}1$ both have a rather
strong attraction.  However, when channel coupling is taken into
account, the attraction is greatly reduced for the $J=1$ channel,
whereas the $J=2$ channel is almost unaffected.  Although they consist
of octet- and decuplet-baryons, their energies are lower than the
$\Lambda \Xi \pi$ three body decay channels. A dibaryon may form in the
state $YIJ=-1\frac{1}{2}2$.

\section{Conclusion}

Taking into account both color screening and quark delocalization, a
systematic study of the effective baryon-baryon potentials has been
performed by using a quark cluster model. The model parameters were
fixed by baryon properties and the color screening parameter was taken
from lattice gauge QCD calculations for linear confinement (or from
$N$-$N$ scattering data for quadratic confinement). Consequently, the
calculated baryon-baryon potentials can be viewed as a parameter free
result. As a general trend, we find that a very strong attraction exists
between two decuplet-baryons. For decuplet and octet baryons, one has a
mild attraction. Very weak attraction, or even repulsion appears in the
octet-octet baryon system. 

Because the "predicted" $N$-$N$, $N$-$\Lambda$ and $N$-$\Sigma$
effective potentials fit the scattering data reasonably well, we expect
that the other baryon-baryon effective potentials are reasonable
predictions as well. At least for those cases for which one has neither
any experimental data nor good theoretical models, our model results
should be useful intermediate guides for studies of such exotics as
mutliquark states, strangelets and of the equation of state of strongly
interacting matter in relativistic heavy ion collisions, where
multistrange baryons constitute a significant component.

For a dibaryon search, these results suggest that experiments should
focus on octet-octet systems with low spin and decuplet-decuplet
systems with high spin.  For octet-octet systems, there are dibaryon
candidates which are loosely bound but stable against strong
interaction decay and have small quark delocalization. This is the
nuclear type of dibaryon and the deuteron is a typical example. For
decuplet-decuplet systems, the most striking resonances are states with
high spin and large quark delocalization. These are a quark matter type
of dibaryon and the $d^{*}$ is a typical example.

Strangeness conserving transition potentials (e.g., $N\Lambda
\rightarrow N\Sigma$) can also be obtained in our model.  We defer
discussion of those results to a later paper.

We thank R. Timmermans for a useful conversation.  This research is
supported in part by the Department of Energy under contract
W-7405-ENG-36 and in part by the National Science Foundation of China.

\pagebreak

\begin{center}
{\large {\bf FIGURE CAPTIONS}}
\end{center}

\noindent Fig.\ 1a ~~ Effective potential in MeV vs.~baryon separation
in $fm$ for N-N channels with\\ 
\hspace*{1.0in} quadratic confinement.

\noindent Fig.\ 1b ~~ Same as in Fig. 1a, but with linear
confinement.

\noindent Fig.\ 2 ~~~ Effective potential in MeV vs.~baryon separation
in $fm$ for $N$-$\Sigma$ channels with\\
\hspace*{1.0in} and without channel coupling.

\noindent Fig.\ 3 ~~~ Effective potential in MeV vs.~baryon separation
in $fm$ for $N$-$\Lambda$ channels with\\
\hspace*{1.0in} and without channel coupling.

\noindent Fig.\ 4 ~~~ Effective potential in MeV vs.~baryon separation
in $fm$ for various forbidden states.

\noindent Fig.\ 5 ~~~ Effective potential in MeV vs.~baryon separation
in $fm$ for $\Delta$-$\Delta$ channels.

\noindent Fig.\ 6 ~~~ Effective potential in MeV vs.~baryon separation
in $fm$ for $YIJ = 000$ channels\\
\hspace*{1.0in} (i.e., with two strange quarks) with and without channel coupling.

\noindent Fig.\ 7 ~~~ Effective potential in MeV vs.~baryon separation
in $fm$ for $N$-$\Omega$ channels with\\
\hspace*{1.0in} and without channel coupling.

\noindent Fig.\ 8 ~~~ Effective potential in MeV vs.~baryon separation
in $fm$ for $\Sigma\Sigma, \Xi\Xi$ and $\Omega\Omega$\\ 
\hspace*{1.0in} channels.

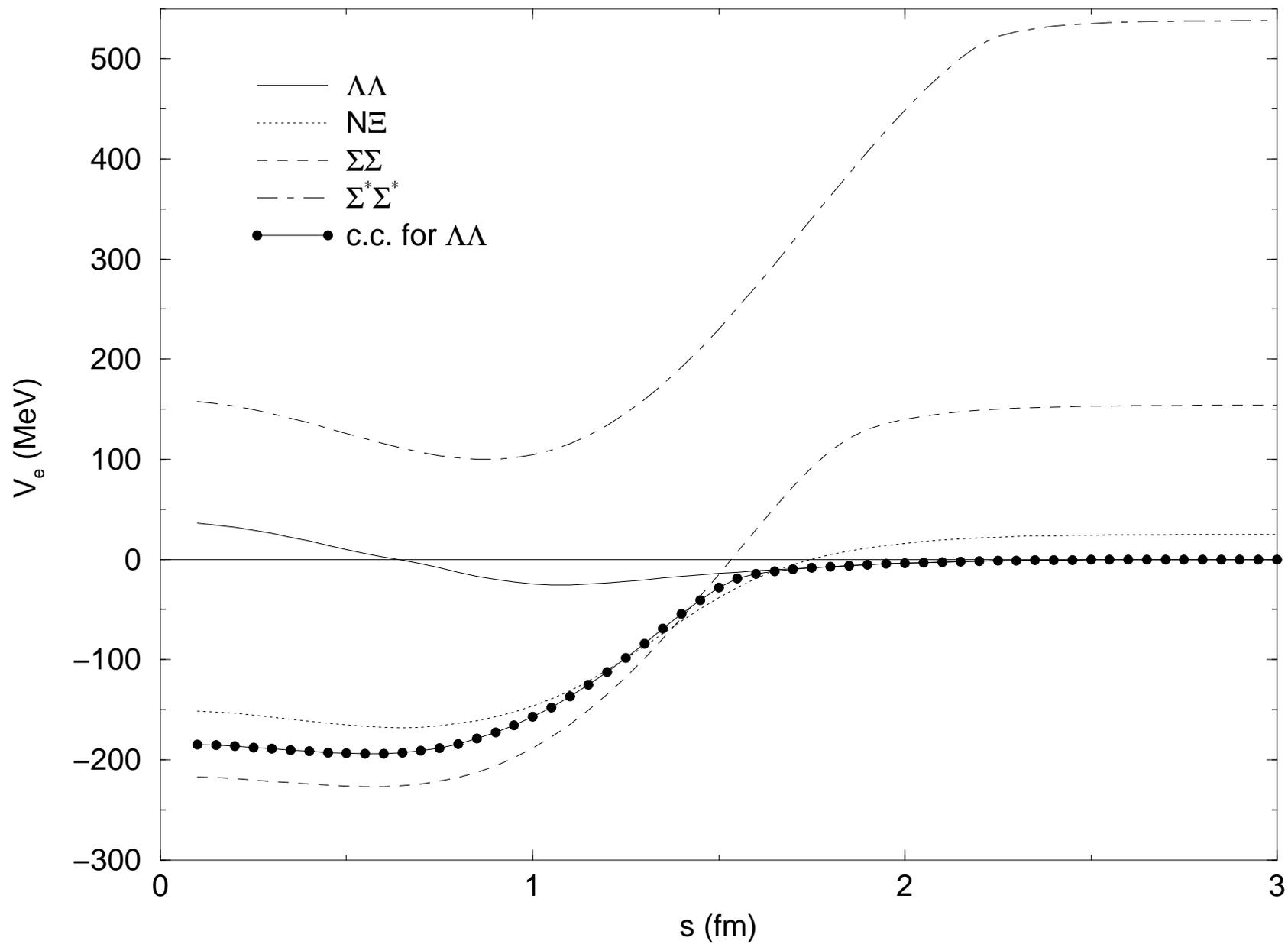
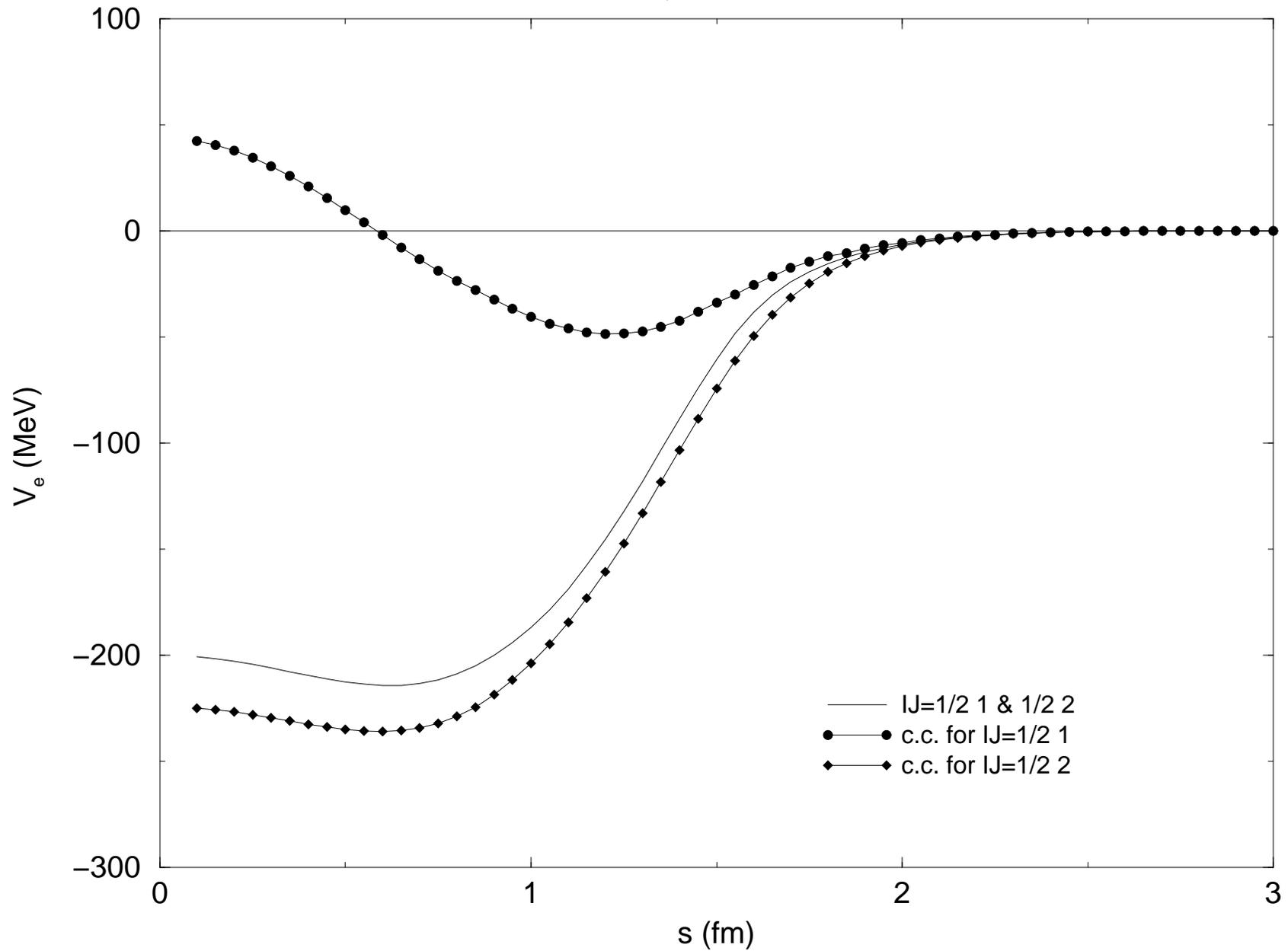
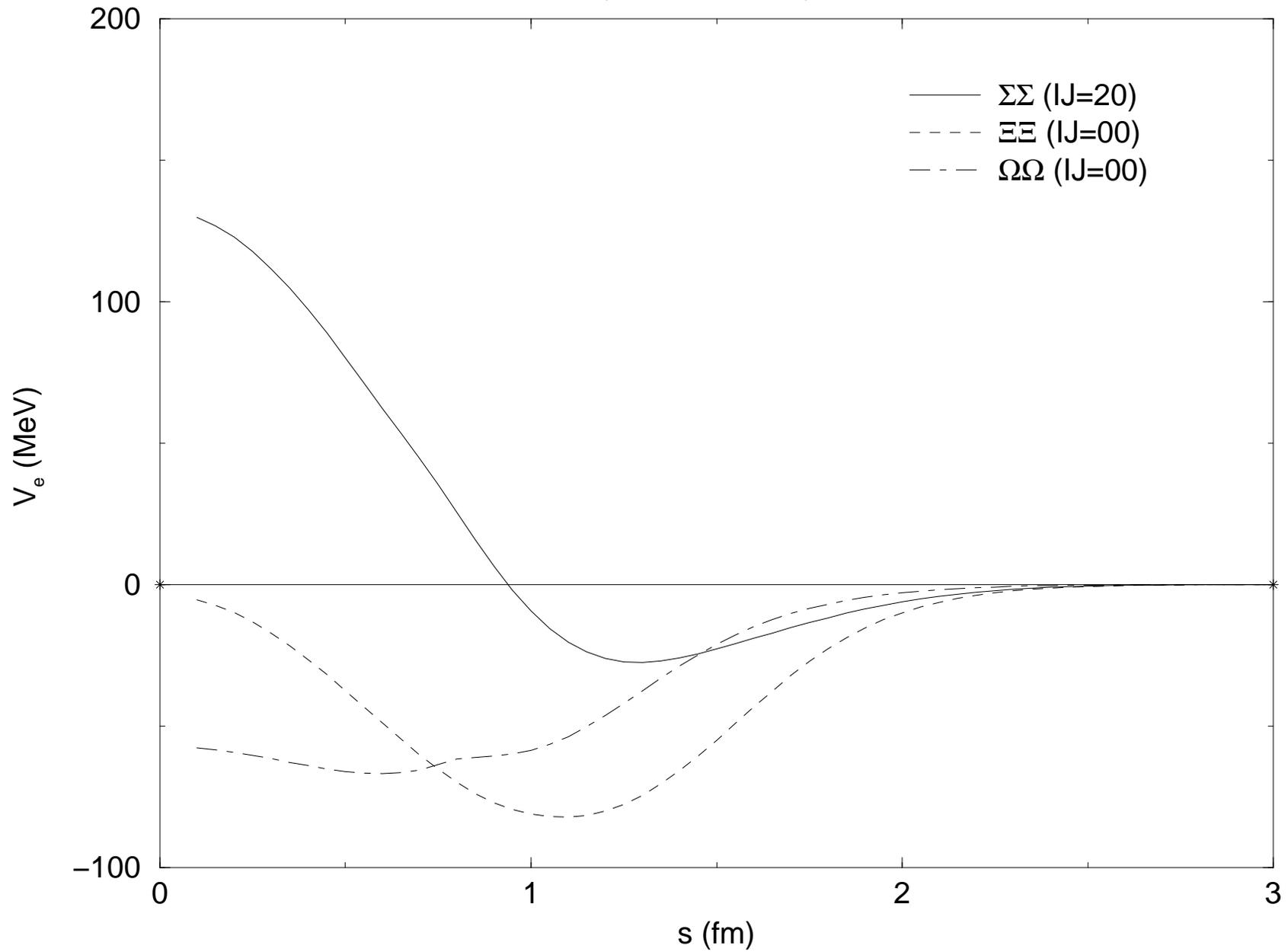
\end{document}